\documentclass[aps,prx,preprint,superscriptaddress]{revtex4-2}

\usepackage[utf8]{inputenc} 
\usepackage{graphicx}
\usepackage{microtype}      
\usepackage{amsmath, amssymb, amsfonts}
\usepackage[hidelinks]{hyperref}
\usepackage[nameinlink]{cleveref} 
\crefname{equation}{Eq.}{Eqs.}
\crefname{section}{Section}{Sections} 
\usepackage{bm} 
\usepackage{mathtools}

\usepackage[T1]{fontenc}
\usepackage{mlmodern}

\newlength{\wid}

\newcommand{\sub}[2]{\ensuremath{#1_{\mathrm{#2}}}}
\DeclareMathOperator{\sat}{sat}

\begin{document}

\title{Modular memristor model with synaptic-like plasticity and~volatile~memory}

\author{Daniel Habart} 
\email[Contact author: ]{daniel.habart@gmail.com}
\thanks{These authors contributed equally to this work.}
\affiliation{Institute of Physiology of the Czech Academy of Sciences, 
Videnska 1083, 14200 Prague 4, Czech Republic}

\author{Stephen H. Foulger}
\affiliation{Center for Optical Materials Science and Engineering Technologies (COMSET),
Clemson University, Clemson, SC 29634 USA}
\affiliation{Department of Materials Science and Engineering, Clemson University,
Clemson, SC 29634 USA}
\affiliation{Department of Bioengineering, Clemson University,
Clemson, SC 29634 USA}

\author{Kristyna Kovacova} 
\affiliation{Institute of Physiology of the Czech Academy of Sciences, 
Videnska 1083, 14200 Prague 4, Czech Republic}

\author{Ambika~Pandey} 
\affiliation{Institute of Macromolecular Chemistry, Czech Academy of Sciences,
Heyrovskeho nam. 2, 16206, Prague 6, Czech Republic}

\author{Yadu R. Panthi} 
\affiliation{Institute of Macromolecular Chemistry, Czech Academy of Sciences,
Heyrovskeho nam. 2, 16206, Prague 6, Czech Republic}

\author{Jiří Pfleger} 
\affiliation{Institute of Macromolecular Chemistry, Czech Academy of Sciences,
Heyrovskeho nam. 2, 16206, Prague 6, Czech Republic}

\author{Jarmila Vilčáková}
\affiliation{Department of Physics and Materials Engineering, Faculty of Technology, Tomas Bata
University in Zlin, Tr. T. Bati 5678, 760 01 Zlin, Czech Republic}

\author{Lubomir~Kostal} 
\email[Contact author: ]{lubomir.kostal@fgu.cas.cz}
\thanks{These authors contributed equally to this work.}
\affiliation{Institute of Physiology of the Czech Academy of Sciences, 
Videnska 1083, 14200 Prague 4, Czech Republic}

\begin{abstract}
Compact models of memristors are essential for simulating large-scale
neuromorphic systems, yet they often do not include description of
complex dynamics like volatile relaxation and synaptic plasticity.  We
introduce a modular, computationally efficient memristor model that
bridges this gap by integrating principles from  physics and
computational neuroscience.  Starting from standard memristive system dynamics, the framework 
incorporates synaptic-like plasticity dynamics, a mapping 
from state variables to cumulative conductance, a volatility module,
and a saturation module.
The plasticity component is inspired by a biological rule for
spike-timing-dependent plasticity (STDP) and is compatible with
the general memristive systems formalism. Finally, we propose a
Laplace transform-based technique to derive the precise form of the mapping from
state variables to cumulative conductance, replacing ad hoc voltage-current relationships with principled construction.

We quantitatively validate the complete model against a comprehensive set of
experimental data from polymeric memristors exhibiting potentiation,
synaptic-like plasticity and volatile decay.  Our work presents a new
paradigm for memristor modeling that is both practical for large-scale
simulation and rich in explanatory power, providing a principled tool
for the design of next-generation neuromorphic hardware.
\end{abstract}


\maketitle


\noindent
\emph{Teaser:} Kernel-based volatility and eligibility-trace plasticity unify short-term memory and learning in polymer memristor compact models.

\section*{Introduction}

Neuromorphic computing allows physical devices to bring memory and
computation closer together, targeting improvements in energy efficiency
compared to traditional von Neumann architectures \citep{r:yang13,
r:ielmini18}. Memristors are one-port devices whose conductance depends
on the history of electrical stimulation \citep{r:chua71, r:chua76} and are
particularly well suited to serve as artificial synapses \citep{r:jo10}.
Their inherent memory, analog tunability, and scalability make them
attractive for both fundamental studies and applications in neuromorphic
hardware \citep{r:yang13}.

Compact models such as TEAM and VTEAM \citep{r:kvatinsky13,
r:kvatinsky15} provide versatile descriptions of memristive switching
\citep{r:wu13prx}, widely used in circuit-level simulations.  
However additional properties must be accounted for in a quantitative and flexible manner for applications in neural networks, such as conductance volatility supporting short-term memory and fading dynamics \citep{r:wang17, r:kim22, r:wang20} and synaptic-like plasticity enabling learning and
long-term adaptation \citep{r:jo10, r:zamarreno11, r:serrano13}.
The recently proposed V-VTEAM model
\citep{r:patni2024} addresses volatility by modifying the
dynamics of the memristive system, but a unified framework that combines synaptic-like plasticity, volatility and saturation while remaining modular and computationally efficient is still lacking.

We propose a \emph{modular memristor model} that augments the standard memristive system description with modules for plasticity, volatility and saturation. This framework unifies:
\begin{enumerate}
  \item a standard memristive systems component utilizing established voltage-driven models
  \citep{r:strukov08, r:kvatinsky15};
  
  \item a novel synaptic-like plasticity component based on local-variable eligibility
  traces \citep{r:gerstner14}, yielding biologically inspired causal and
  anti-causal weight updates;

  \item a cumulative conductance function that maps state variables to conductance;

  \item a volatility module inspired by linear viscoelasticity \citep{r:coleman61,
r:larson99}, where material response depends on a hereditary convolution kernel; and
  
  \item a saturation module implemented as a linear-nonlinear scheme in conjunction with volatility, to produce bounded device conductance.
\end{enumerate}

One can select the memristive core, cumulative conductance mapping, decay kernel and saturation function to match a given device and enable or disable volatility and/or STDP without changing the underlying modeling framework. 
Although we present the framework in voltage-driven form consistent with common experimental protocols, its modular structure also permits straightforward translation to current-driven variants. 


We validate the approach using experimental data from polymeric memristors developed by Pfleger,
Foulger, and co-workers \citep{r:pfleger23, r:pfleger24, r:foulger21,
r:foulger24}.  
These memristors consist of a thin film of
poly(N-(3-(9H-carbazol-9-yl)propyl)methacrylamide) sandwiched between
ITO and top metallic electrodes. 
The devices studied exhibited pronounced
hysteresis in their I-V characteristics, bistable resistive switching effects at higher voltages of +/-5 V (with a maximum ON/OFF
ratio of around 200 and retention time exceeding 2 hours)\citep{r:pfleger23}, and analog, pulse-driven conductance modulation at lower voltages that resembles potentiation and depression of neuronal synapses
\citep{r:pfleger24}.
For this class of devices, the internal conductance state depends on the
history of applied voltage via a convolution with a kernel that decays approximately as \(1/t\).  
This form reflects a broad distribution of relaxation times, consistent with
disordered systems and percolation phenomena in which conduction
pathways evolve stochastically over time \citep{r:stauffer94, r:song23}. 


Using an appropriate measurement protocol, the distinct effects of the individual components may be effectively separated.
To this end, we developed a measurement procedure that exploits the modular structure of the framework to efficiently estimate the model parameters, enabling near-independent optimization of each functional component.

Finally, we introduce a technique based on the Laplace transform
to determine the functional form of the cumulative conductance mapping
for a given device.
Owing to the simplicity of the model framework, this form may provide useful
insight to the characteristic macroscopic behavior of the device.

The model reproduces both volatile memory and synaptic plasticity observed in polymeric memristors, while maintaining computational efficiency for large-scale neuromorphic simulations.
We believe this approach will help bridge the gap between device-level
physics and system-level neural computation, with particular relevance
for sustainable, polymer-based memristive hardware.

\section*{Results}
We present the theoretical model and its application to hardware for parameter fitting. 

\subsection*{Model}\label{sub:model}
For reference, we recall the general framework for voltage-controlled,
time-invariant one-port memristive systems introduced by
\citet{r:chua76}. Such a device is described by a differential equation
governing the evolution of a vector of internal state variables $\bm x$,
together with an algebraic equation relating the current and voltage
through a conductance function $G(\bm x)$:
\begin{equation}
    \begin{aligned}
    i &= G (\bm x) \, v  \\
    \dot{\bm x} &= f(\bm x, v) \, .
    \end{aligned} \label{eq:chua}
\end{equation}
Here $v$ is the voltage across the device, $i$ is the current through
the device and the functions $f$ and $G$ are such that we assume a
unique solution to the equation. Throughout the following, the symbol
$\,\dot\, \,$ always denotes differentiation with respect to time.

Our approach is inspired by the memristive system framework, but differs
from it in several important aspects.  Following \cref{eq:chua}, we
introduce a vector of four internal state variables: $s,\, w, \, x,
\,y$, along with a description of their respective dynamics.  
To incorporate STDP into our memristor model, we introduce a formulation of spike timing-dependent plasticity dynamics  (variables $w,\, x$ and $y$) for one-port electronic devices, compatible with the standard formulation of memristive systems.

\begin{figure*}
\centering
\includegraphics[width=\linewidth, height= \textheight, keepaspectratio ]{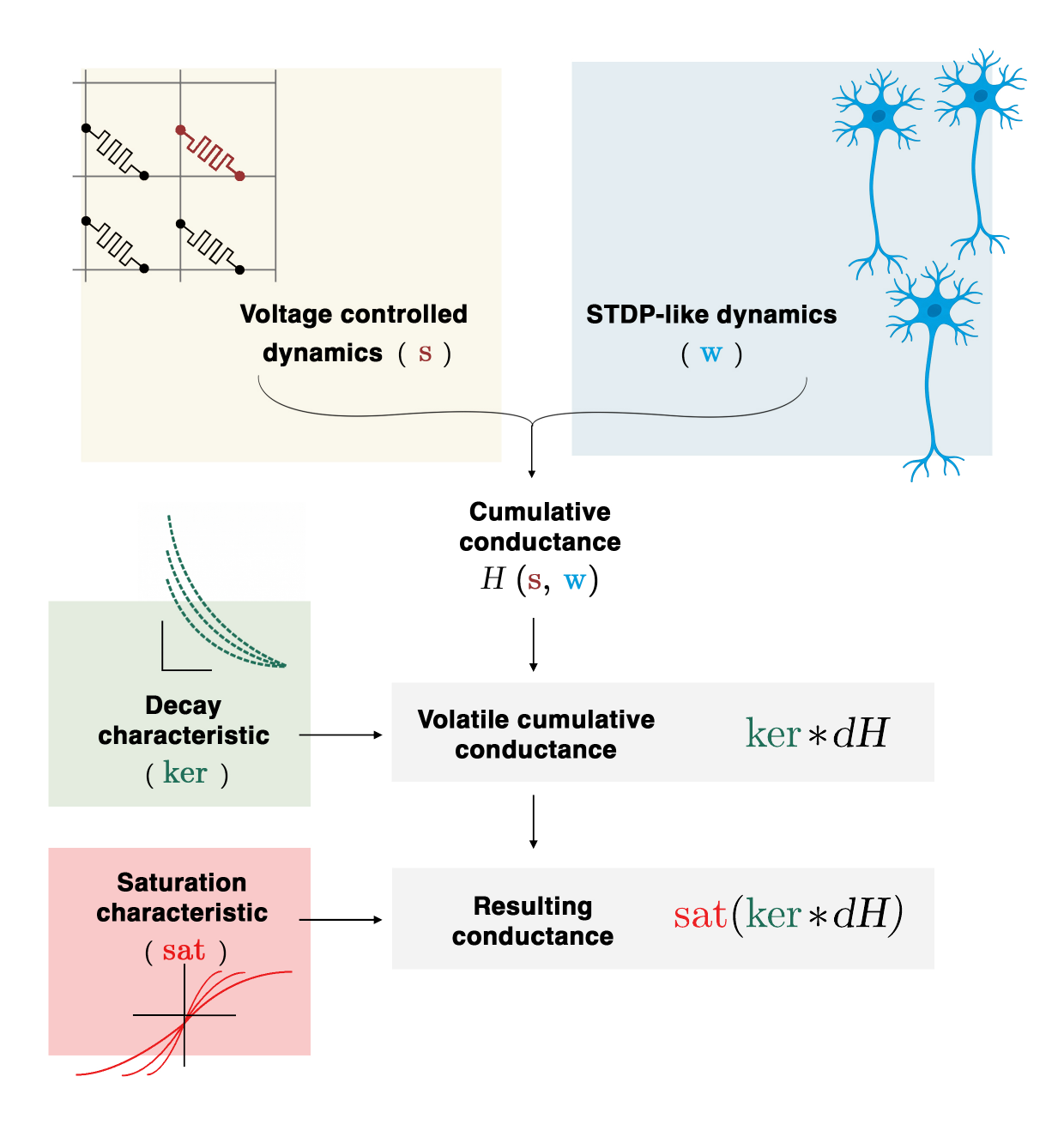}
\caption{Schematic representation of the model.
The model combines two independent components into the cumulative conductance $H(s,w)$: voltage-controlled switching dynamics providing the memristive core (state variable $s$) , and STDP-like synaptic plasticity implementing learning rules via eligibility traces (state variable $w$). 
The cumulative conductance then undergoes volatile decay through viscoelastic-inspired convolution with a decay kernel $\ker()$ and 
the resulting device conductance $G$ emerges after a saturation nonlinearity. 
}
\label{fig:schematic}
\end{figure*}

However,
unlike in the formulation above, our device conductance function $G$ is not directly a function of the state parameters. 
Instead, we introduce a kernel function $\ker(\,)$ to capture the
temporal decay  of conductance, a "cumulative conductance" mapping
$H(s,w)$ from state variables to non-volatile device conductance and a
saturation function $\sat(\,)$ to model device conductance saturation.  
A schematic of the modules' relationships is provided in \Cref{fig:schematic}.



\subsubsection*{Memristive system dynamics} 
This component builds upon established voltage-driven memristive models, suitable as the basis for non-volatile behavior. A simple and common possibility for the memristive model core is the linear voltage controlled memristor model \citep{r:strukov08}:
\begin{equation}
    \dot s = \mu \, v\, . \label{sdot:muv}
\end{equation}

A more detailed approach due to \citep{r:kvatinsky15} is also a popular choice.

Notably, we have omitted the windowing functions in the above equation,
because in our approach the state parameters are unbounded at this stage
of the modeling process.  Hence the so-called linear
memristor model of \cref{sdot:muv} is truly a linear differential
equation at this stage. 
We resolve non-physical
behavior by the saturation component of the model. 


It should be emphasized that any kind of voltage-controlled memristor
state-variable differential equation in the form specified in
\cref{eq:chua} is compatible with our framework.

\subsubsection*{Adaptation of STDP to a memristive device}
In this section, we adapt the local variables synaptic STDP model, as
presented in \citep{r:gerstner14}, to a one-port physical electronic
device, making it compatible with the standard memristive systems framework.
The model provides a phenomenological description of synaptic weight
parameter changes based on  relative timing of two classes of spiking
inputs: presynaptic and postsynaptic.  Notably, due to its abstract
formulation, it can be generalized beyond neural systems,
provided the two input classes are reinterpreted in an appropriate
manner. 

To begin, recall the synaptic model's original formulation and
interpretation. We consider the synaptic weight variable $w$, whose
dynamics are driven by presynaptic and postsynaptic spike trains
$\sub{S}{pre}$ and $\sub{S}{post}$, respectively.  Denote by $(t_f)$ and
$(t^n)$ the presynaptic and postsynaptic spike time sequences.  The
spike trains are then modeled as a series of Dirac impulses at the
corresponding times: \[
    \begin{aligned}
        \sub{S}{pre}(t) = \sum_f\delta(t-t_f), \quad
        &\text{and}\quad
        \sub{S}{post}(t) = \sum_n\delta(t-t^n).
    \end{aligned}
    \]

Define two auxiliary variables $x$ and $y$, representing the presynaptic
and postsynaptic traces, which model the capacity for change in the
weight variable $w$. They are incremented by their respective spike
trains and decay exponentially between spikes:
    \begin{align*}
            \dot x(t) &= -\frac{x(t)}{\sub{\tau}{+}}+ \sub{S}{pre}(t)\\
            \dot y(t) &= -\frac{y(t)}{\sub{\tau}{-}} +  \sub{S}{post}(t),
    \end{align*} where $\sub{\tau}{+}, \, \sub{\tau}{-}$ are time constants.
    
Finally, the weight variable $w$ is updated at postsynaptic spike times
proportionally to the presynaptic trace $x$, and negatively at
postsynaptic spike times in proportion to the postsynaptic trace $y$:
    \begin{align*}
        \dot w(t) &=  x(t)\,\sub{A}{+}(w)\,\sub{S}{post}(t) -y(t)\,\sub{A}{-}(w)\,\sub{S}{pre}(t) \,.
    \end{align*}
The functions $\sub{A}{+},\, \sub{A}{-}$ are weight dependent scaling factors.

Our aim is to reformulate this model for a one-port device, which differs from biological synapses in several important respects.  
The key idea is to
replace discrete spike trains with continuous voltage signals, while
interpreting one terminal as the 'presynaptic' terminal and the other as
the 'postsynaptic' terminal.  With this identification, the respective
spike trains correspond to positive and negative voltages across the
device.  Accordingly, we substitute the spike trains at each synaptic
terminal with the appropriate signed voltage function, as illustrated
below:
    \begin{align*} 
    \sub{S}{pre}(t) &\implies v^+(t)\\
    \sub{S}{post}(t) &\implies v^-(t)\, ,
    \end{align*}
where $v^+,\, v^-$ are the positive and negative parts of $v$,
respectively, with the convention $v = v^+-v^-$.

Adding variables $w,\, x,\ y$ as additional state parameters of
the memristive device model, we can formulate the synaptic-like plasticity component for the one-port device as
follows:
\begin{equation}
\begin{aligned}
    \dot x &= -\frac{x}{\sub{\tau}{+}}+ v^+\\
    \dot y &= -\frac{y}{\sub{\tau}{-}} + v^- \\
    \dot w &=   x \, \sub{A}{+} (w)\, v^- -y \, \sub{A}{-}(w)\, v^+\,,
\end{aligned}\label{stdpw}
\end{equation} where $\sub{\tau}{+}, \, \sub{\tau}{-}$ are time constants and the functions $\sub{A}{+},\, \sub{A}{-}$ are weight dependent scaling factors.
We denote by  $v^+,\, v^-$ the positive and negative parts of $v$,
respectively, with the convention $v = v^+-v^-$.

We differ from \citep{r:gerstner14} in our approach by considering
unbounded $w$ at this stage.  
As a result the functions $A_+,\, A_-$ do
not need to contain a windowing function to force bounds onto the weight
parameter. 
We note that in the model, the memristive system components may incorporate coupling by allowing $\sub{A}{+}$ and $\sub{A}{-}$ to depend on $s$.

\subsubsection*{Volatility kernel and cumulative conductance}
In order to model memristor volatility characteristics, we implement a
standard convolution-based approach, inspired by linear viscoelasticity
theory \citep{r:coleman61}.  A key advantage is that this approach
enables the use of $1/t^\alpha$ volatility kernels that are otherwise
computationally unavailable.
We select a suitable kernel function $\ker()$,
which characterizes the device's conductance volatility, and the
\emph{cumulative conductance function} $H(s,w)$, which describes
nonvolatile behavior of conductance.  
The resulting \emph{volatile
cumulative conductance} $ \sub{H}{vol} $ is calculated as the
convolution
\begin{equation}
    \sub{H}{vol} = \ker \ast \,dH(s,w).\label{g:decay}
\end{equation}

Note that the convolution is calculated with respect to time, since the
model state variables $s$ and $w$ are time-dependent.  Similarly, the
differential $dH$ must also be taken with respect to time, written
explicitly: 
\begin{equation} \label{eq:dhdif} dH = \partial_s H(s,w)\,\dot s +
\partial_w H(s,w)\,\dot w\,.  
\end{equation}

Regarding the kernel function, we may distinguish two cases, based on
its limiting behavior. 
For example, if the kernel is the Heaviside
function, no volatile behavior is present (pure ``elastic'' regime),
and~\cref{g:decay} simplifies to \[ \sub{H}{vol} = H(s,w).  \] 
This relation motivates the label "cumulative conductance" for the function $H$; it captures the functional relationship of state variables and device conductance in the non-volatile case.  
Conversely, if
$\displaystyle \lim_{t\rightarrow \infty} \ker(t) = 0$, we are in the
viscoelastic regime, where the system decays to its equilibrium over
time.
Our model is therefore sufficiently general to describe devices
exhibiting a combination of volatile and non-volatile characteristics.

The form of the kernel and cumulative conductance functions is device-specific and also depends on the chosen voltage-controlled model. Inspired by the presentation in \citet{r:kvatinsky15}, we provide ad-hoc variants for the kernel function and for the cumulative conductance function. 

First, a linear and exponential relationship of the cumulative conductance on the state variables are presented:
\begin{align*}
    dH(s,w) &= h_1\,\dot s +  h_0\,\dot w \\
    dH(s,w) &= h_1\,e^{a\,s}\, \dot s + h_0\,e^{b\,w}\, \dot w\, .
\end{align*}
The terminology corresponds to the usual relationships in the case of the purely elastic regime ($\ker()$ is the Heaviside function), as discussed above. Thus both versions generalize the current-voltage relationships presented in \citet{r:kvatinsky15}.

Similarly, one can consider variants for the kernel:
\begin{align*}
 \ker(t) &= e^{-at}\\
 \ker(t) &= \frac{1}{(t+\varepsilon)^a}\, ,
\end{align*} for $t>0$ and identically zero for negative times. 
The power-law kernel may be offset by a small positive $\varepsilon$ to avoid the singularity at zero.

In addition, we offer a rigorous but data-specific approach for deriving the functional forms of the volatility characteristics $\ker()$ and $H$ in \Cref{eq:kernel} and \Cref{eq:cumulative}, respectively. We do so using data from pulse stimulation of the device, idealized by a single box function. 
Consequently, in the fitting section below we consider a conductance function of the form
\begin{align}\label{eq:dH}
dH(s,w) = (h_{11}\, |s|^a +h_{10} )\dot{s} + (h_{01}|w|^a+h_{00})\dot w\,.
\end{align}

As a final remark, we note that convolving the kernel with the derivative $dH$ is useful because it allows the convolution term to be effectively 'reset' by the state parameter derivatives $\dot s,\, \dot w$, which is necessary for the kernel-based approach to work correctly.

\subsubsection*{Saturation and resulting conductance} 
Since the approach described so far allows
memristor conductance values that are unbounded and could even become
negative, it is necessary to limit the conductance to realistic bounds.
To obtain the resulting conductance from the volatile, unbounded version
\cref{g:decay}, we employ the linear-nonlinear scheme
\citep{r:gerstner14}; a standard method to introduce nonlinearities,
such as saturation, into a linear system.

Hence, we calculate the resulting conductance $G$ as:

\begin{equation}
    G = \sat (\sub{H}{vol})\, ,\label{g:sat}
\end{equation} where $\sat(\,)$ is a chosen nonlinear saturation function.

Sigmoid functions are a common choice for saturation functions given
their flexibility, hence here we adopt the logistic function:
\begin{equation}
    \sat (h) = (\sub{g}{max}-\sub{g}{min})\frac{1}{1+e^{-A\,h}}+\sub{g}{min}\,,
\end{equation}
where
\[
    A=\frac{4}{\sub{h}{max} - \sub{h}{min}}
\]
to best align the linear portion of the logistic function with the range of conductance.







\subsection*{Fitting to experimental data}

Application of our model to experiments relies on comprehensive
electrical characterization data, using protocols designed to isolate
different aspects of memristive behavior. Measurements were done using
trains of stimulation pulses (±500\,mV, 20\,ms duration) with
interleaved read pulses (-50\,mV, 50\,ms duration) to monitor
conductance evolution.  Owing to observed voltage-threshold behavior,
the contribution of the read pulses was considered negligible and
therefore omitted from the model simulations.

Two measurement protocols formed the empirical basis for model
development.  First, relaxation measurements following stimulation at
different intensities revealed volatile memory dynamics, with
conductance monitored over timescales from tens of milliseconds to
hundreds of seconds.  Second, specialized timing protocols captured STDP
characteristics by varying the delay between paired pulses, mapping the
full spectrum of causal and anti-causal plasticity.

To derive the model for a given device, we first extract the
characteristic volatility kernel from the relaxation measurements using
an approximation technique. We then apply the Laplace method described
below to determine the form of the cumulative function and
estimate the corresponding parameters.  Next, we estimate the parameters
of the STDP module from measurements and finally set the saturation
module coefficients  either to minimally affect the model, or by
performing a joint optimization of all parameters.

\subsubsection*{Determining the kernel-function form}
\begin{figure}[t]

    \centering
    \includegraphics[width=85mm, keepaspectratio]{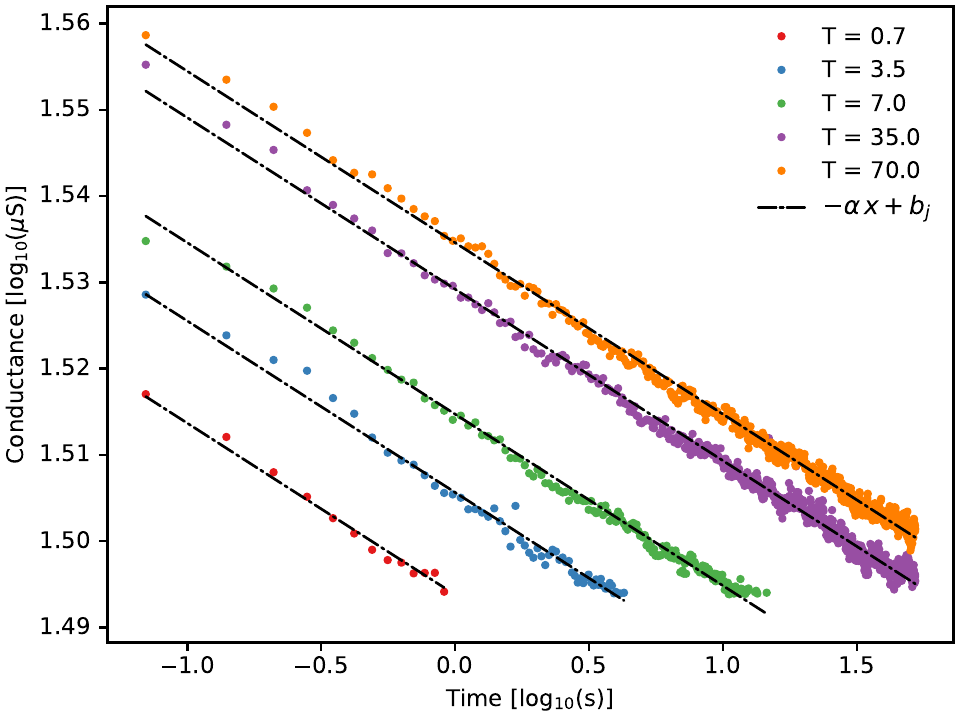}
    \caption{Log–log representation of post-potentiation conductance relaxation in the PCaPMA memristor. Each curve shows the decay of conductance after a train of trigger pulses with different total stimulation time $T_j$. When plotted in double-logarithmic coordinates, the initial segments of all traces are close to straight lines over more than two decades in time, indicating a power-law-like decay and ruling out a simple exponential relaxation. 
    }
    \label{fig:decay-log}
    \end{figure}

In this section we derive the functional form of the kernel from conductance relaxation measurements.
The decay segments of the conductance evolution are shown
in log-log coordinates in \Cref{fig:decay-log}, with data that deviate
from the linear behavior omitted.  It is visible that the initial slopes
of the decay are linear up to over two orders of magnitude, which
excludes exponential decay.  We fit the data using functions of the form
\(-\alpha x+b_j \), where \( \alpha  \) is the common slope and \( b_j
\)  is the individual offset.

The function \(-\alpha x+b\) is transformed into lin-lin coordinates as 
\[
\bar{b} \, \frac{1}{x^\alpha}\, ,
\]
for a suitably modified \( \bar{b}\). 
The expression is undefined for \(x=0\), so a small positive shift by $\varepsilon$ is introduced.
The addition of \(\varepsilon\) preserves the short-time characteristic of the relaxation, but avoids the singularity at zero.

Owing to the power-law characteristic, we are considering the kernel
function in the form
\begin{equation}
    \ker (t) = 
    \begin{dcases}
        \makebox[\widthof{$\dfrac{1}{(t + \varepsilon)^{\alpha+1}}\, ,$}][c]{$0\, ,$} & \text{if } t < 0 \\
        \dfrac{1}{(t + \varepsilon)^{\alpha+1}}\, , & \text{if } t \geq 0\,,
    \end{dcases}\label{eq:kernel}
\end{equation} for $\alpha > 0$.
To ensure proper scaling, the kernel must be normalized. To this end set 
\[
I_\varepsilon = \int^{\infty}_{\varepsilon} \frac{1}{x^{\alpha+1}}\, dx \,  = \frac{1}{\alpha\cdot \varepsilon^\alpha},
\]
and define the resulting normalized kernel as 
\[
\frac{\ker (t)}{I_\varepsilon}.
\]

\subsubsection*{Finding the kernel-function coefficient $\alpha$}
\begin{figure}[t]

    \centering
    \includegraphics[width=85mm, keepaspectratio]{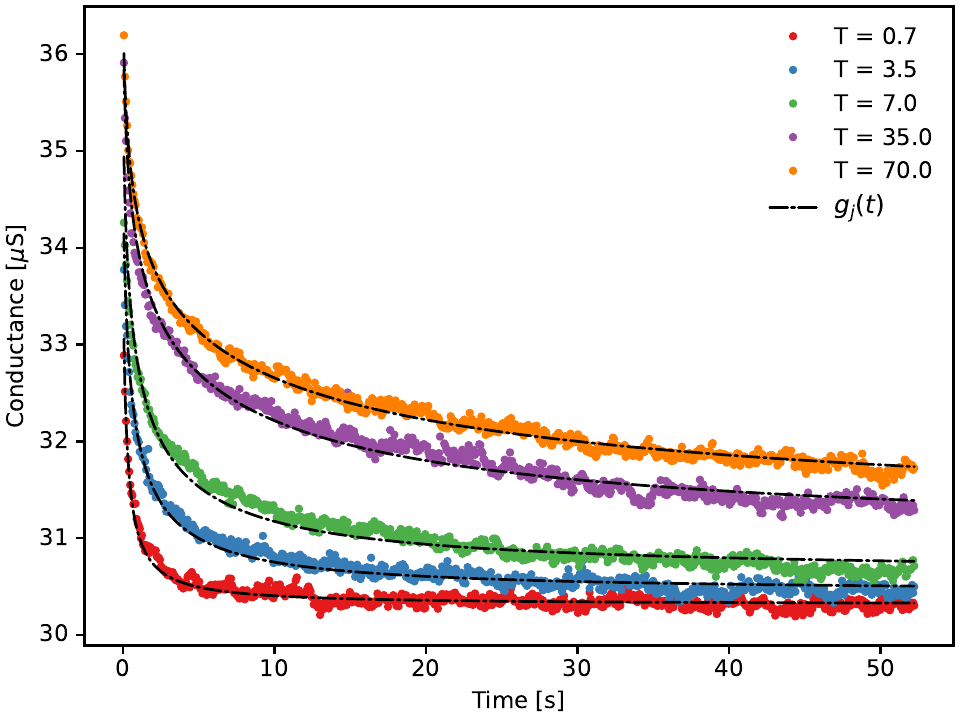}
    \caption{Kernel function coefficient estimation. Each trace corresponds to stimulation by pulses over the duration $T_j$. For each trace, a fit $g_j(t) = b_j \, (\frac{1}{(t+\varepsilon)^\alpha}- \frac{1}{(t+\varepsilon + T_j)^\alpha}) + c_j$ is performed using two trace-specific parameters $b_j$ and $c_j$ together with the global coefficient $\alpha$ to be estimated. }
    \label{fig:decay}
    \end{figure}

To estimate the value of $\alpha$ that results in the measured
conductance $G$, assume the conductance data are the response to a box
function of applied voltage $b\,\chi_{(-T,\,0)}$, where $b$ is a scalar
and $T$ is the time of applied stimulation.

\[
    G(t) = b\, \chi_{(-T,\,0)}\ast \ker (t) + c\, ,
    \]
    for a suitable stationary conductance constant $c$.
We calculate
    \begin{align*}
        G(t) &= b \,\int^{t}_{-\infty} \ker (t-\tau )\, \chi_{(-T,\,0)} (\tau) \, d \tau  +c
        =  b\, \int^{0}_{-T}  \ker (t-\tau )\,d \tau +c\\
        &= b\, \int^{T+t}_{t}  \ker (u)\,d u +c = b\,\big(K(t+T)-K(t)\big) +c
    \end{align*}
    where
\begin{align*}
K(t) = \int^{t}_{-\infty}\ker (\tau)\, d\tau =         
\begin{dcases}
    \makebox[\widthof{$1-\dfrac{\varepsilon^\alpha}{(t+\varepsilon)^\alpha}\, ,$}][c]{$0\, ,$} & \text{if } t < 0 \\
    1-\dfrac{\varepsilon^\alpha}{(t+\varepsilon)^\alpha}\, , & \text{if } t \geq 0,
\end{dcases}
\end{align*}
is a primitive function to $\ker() $. 
To derive the decay characteristic after the pulse stimulation from time $-T$ to $0$, fit the decay of the graph to the function for $t \geq 0$
\[
    G(t) =  b\,\big(K(t+T)-K(t)\big) +c =b\, \Big( \frac{\varepsilon^\alpha}{(t+\varepsilon)^\alpha} - \frac{\varepsilon^\alpha}{(t+T+\varepsilon)^\alpha} \Big) + c
    \]
We repeat the above procedure for every measured trace and find an optimal $\alpha$ over all five datasets, as shown in \Cref{fig:decay}.

We have set $\varepsilon$ to the time resolution of the measurements
($70$ms) and calculated the value of alpha to be $\alpha= 0.029$, which
corresponds to the volatility following a power-law with an exponent
$1.029$.

\subsubsection*{Derivation of cumulative conductance}
In this section we describe a method for determining the form of the
cumulative conductance function $H$.  We do so using data from pulse
stimulation, again idealized by a single box function similarly as above.  
The procedure yields only the dependence on the
parameter $s$, the dependence on $w$ must be copied, or derived in a
different way.

Our goal is to find the cumulative conductance function such that the
prediction given by the model results in the traces $g_j$ from
\Cref{fig:decay}. Note that because the following utilizes the Laplace
transform, we must restrict ourselves to positive times, hence in
contrast to the discussion above we work with the single box
function supported on the interval $(0,\,T)$, i.e.\! $\chi_{(0,\,T)}$.

\begin{figure}[t]

        \centering
        \includegraphics[width=85mm, keepaspectratio]{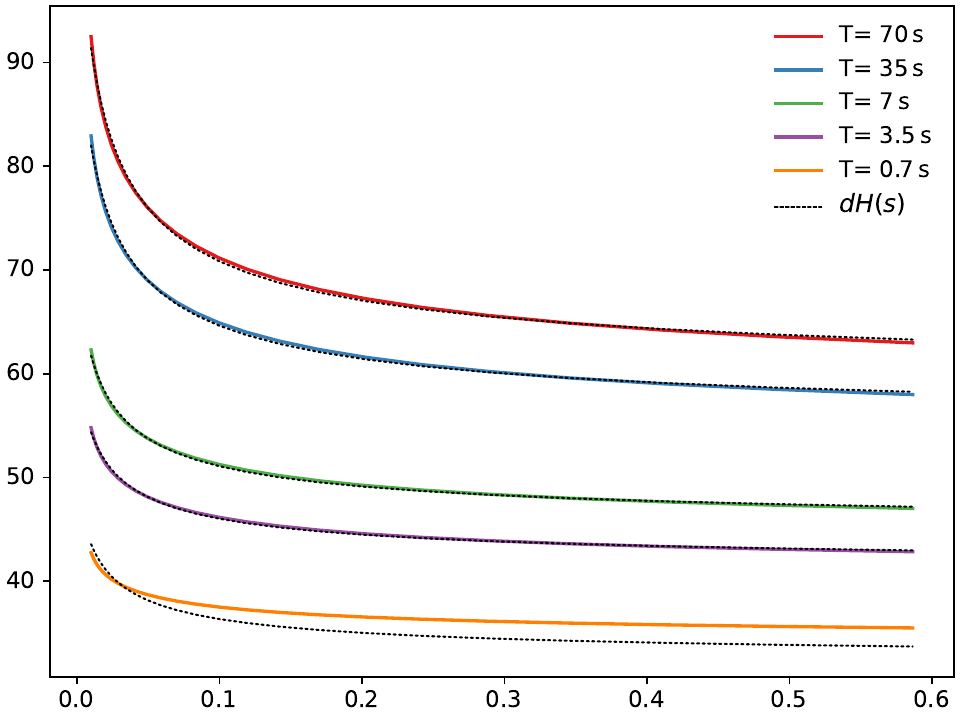}
\caption{Cumulative conductance function estimate. Each curve shows the
estimated dependence of $\partial_s H(s)$ on the state variable $s$,
reconstructed from relaxation traces with different stimulation
durations $T$ using the discussed procedure. The
parameters $h_0$, $h_1$  are chosen so that the traces for all
$T$ collapse onto a power-law-like curve with a common $a = -0.585$, well approximated by the parametric form $dH(s) = (h_1\,|s|^a + h_0)\,\dot s$. This
justifies the simple functional form of $dH$ used in the main text.}
        \label{fig:dH}
        \end{figure}

\begin{enumerate}
    \item We use the previously calculated traces dependent on $T_j$: 
    \begin{align*}
        g_j = b_j\,\ker\ast \chi_{(0,\,T)}+c_j\,.
    \end{align*}
    \item Find explicit functions $b(T)$ and $c(T)$ that describe the dependence of $b_j$ and $c_j$ on $T_j$, respectively. 
    In our case, we find by minimization appropriate constants $b^i$, $c^i$ such that
    \begin{align*}
    b(T) &= b^1\,\log(T+b^2)+b^3\\
    c(T) &= c^1\,\log(T+c^2)+c^3.
    \end{align*}
    Then the traces become 
    \begin{align}
    g = b(T)\, \ker\ast \chi_{(0,\,T)} + c(T)\,.\label{eq:g-trace}
    \end{align}
    \item Neglecting the saturation term, our model predicts $g=\ker\ast dH(s) + g_0$, where $dH(s(t)) = \partial_s H(s(t))\,\dot{s}(t)$. Recall the voltage-control linear differential equation \cref{sdot:muv}, $\dot s = \mu\, v$, and consider a normalized voltage input $v = \chi_{(0,T)}$. Then it holds that $s(t) = \mu\,t$ and by setting \cref{eq:g-trace} equal to our model we obtain:
    \[
    \ker\ast \big(\partial_s H(s)\,\mu\, \chi_{(0,T)}\big)\,  + g_0 = b(T)\,\ker\ast \chi_{(0,\,T)} + c(T)\,.
    \]
    \item We apply the Laplace transformation to both sides and perform some rearranging to derive the form of $dH$, valid for $t\in [0,T]$:
    \begin{align*}
    \mathcal{L}\big(\ker\big) \,\mathcal{L}\big(\partial_s H(s)\,\mu\, \chi_{(0,T)}\big)\,  + \frac{1}{s}g_0 = b(T)\,\mathcal{L}\big(\ker\big) \mathcal{L} \big(\chi_{(0,\,T)}) + \frac{1}{s}c(T)
    \\
    \mu \partial_s H(\mu\,t) = \mathcal{L}^{-1}\Big\{\, \big( b(T)\,\mathcal{L}(\ker)\,\mathcal{L} (\,\chi_{(0,\,T)})+ \frac{1}{s}(c(T)-g_0)\big)/\mathcal{L}(\ker) \,\Big\}\,.
    \end{align*}
    \item In our case, the calculation of the Laplace transform of $\ker()$ and $\chi_{(0,T)}$ is performed, up to normalization constant, as
    \begin{align*}
    \mathcal{L}(\ker)(s) &=  s^\alpha e^{s\varepsilon} \, \Gamma(\alpha,\,s\varepsilon) \\ 
    \mathcal{L}(\,\chi_{(0,T)})(s) &= \frac{1}{s}\,(1-e^{-sT})\, ,
    \end{align*}
    where $\Gamma(a,x) = \int_x^\infty t^{a-1} \,e^{-t}\,dt $ is the upper incomplete Gamma function.

    As a result, the above procedure yields the following expression for positive values of $x$:
    \[
    \partial_s H(x) = h_1\, x^a+h_0\, .
    \]
    In general there may be some dependence of $H$ on $T$ (refer to \Cref{fig:dH}), and we choose the parameters $h_1,\, h_0$ of the cumulative function that minimizes such dependence.
    To obtain $dH$, extend the relationship symmetrically for $x<0$, hence the resulting functional form is 
    \begin{align}\label{eq:cumulative}
        dH(s) = (h_1\, |s|^a + h_0)\, \dot s 
    \end{align}
\end{enumerate}

\subsubsection*{Finding cumulative conductance function coefficients}\label{se:h}
\begin{figure}[h]
\centering
\begin{minipage}[b]{0.4\textwidth}
\includegraphics[width=85mm, keepaspectratio]{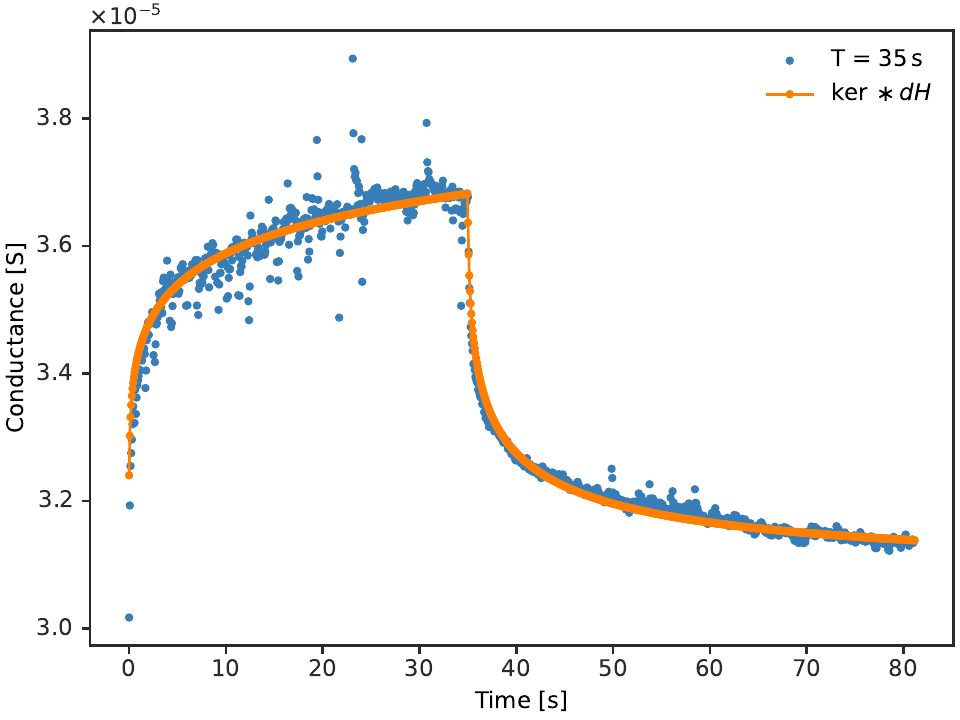}
\end{minipage}

\begin{minipage}[b]{0.4\textwidth}
\includegraphics[width=85mm, keepaspectratio]{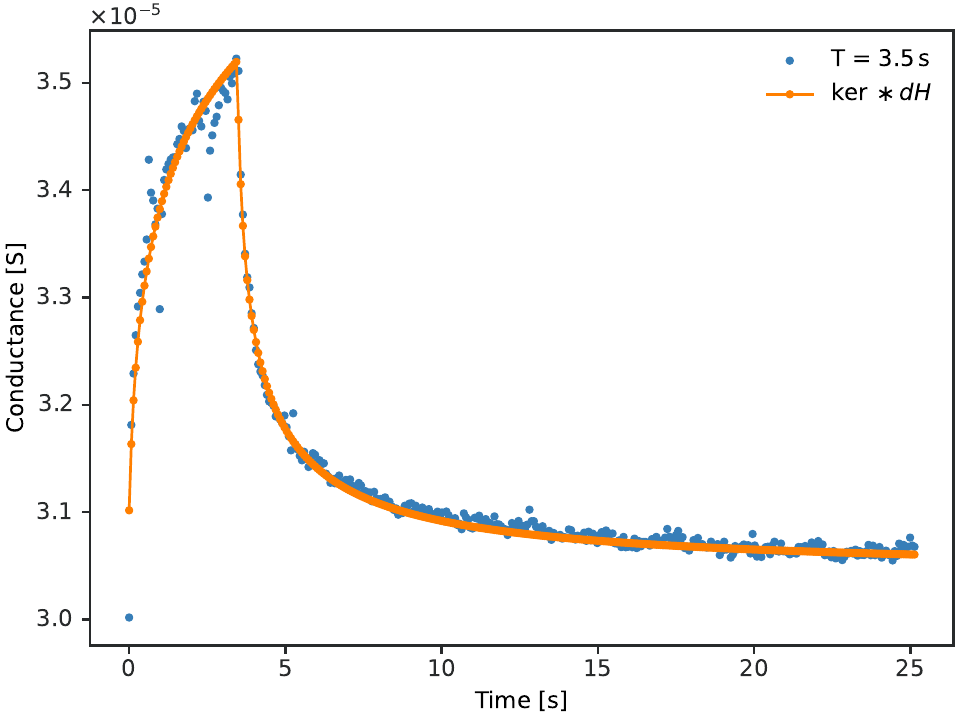}
\end{minipage}

\caption{Comparison between the model and the experimental data for stimulation input of $T=35$\,s and $T=3.5$\,s. 
The measured conductance decays after potentiation by a pulse train of total duration $T$ and the solid line shows the corresponding numerical simulation of the model.
The overall good agreement across the entire decay window indicates that a single, material-specific kernel captures the volatile dynamics for different stimulation protocols.}
\label{fig:dh1}
\end{figure}

Having determined the characteristic of the decay, we now turn to estimating the parameters for the remaining parts of the model.
Neglecting the saturation component for now, the resulting conductance \( G \) is predicted by the model as:
\[
    G =  \ker \ast dH + g_0\,.
\]
We are considering $dH$ in the form derived in \Cref{eq:cumulative}:
\[
dH(s) = (h_1\, |s|^a + h_0)\, \dot{s},
\]
where we already have estimated the optimal coefficient $a=-0.585$, and are looking to estimate the parameters $h_0$ and $h_1$. For demonstration, we perform the fit on the stimulation segment of the traces above, this time with the precise input data. We obtain the graphs in \Cref{fig:dh1}.

\subsubsection*{Modeling STDP}

To estimate the coefficients of the STDP component, we extend the
previously computed part of the model with the dependence on the weight
parameter $w$. Specifically we are now looking to estimate the
coefficients from \cref{stdpw}: $\sub{A}{+},\, \sub{A}{-},\,
\sub{\tau}{+},\, \sub{\tau}{-}\,$, as well as the coefficients from the
STDP component of $dH$; $h_{01}$ and $h_{00}$.

The data under evaluation consists of three conductance measurements per run. 
The first determines the steady-state conductance $g_0$, while the other two quantify the conductance change $\Delta G$: one measured 10ms after the end of the second testing pulse (denoted $t_1$) and another measured after 100ms (denoted $t_2$). 
We optimize the parameters for the first measurement, since the second one should be a
result of the volatility modeled by $\ker()$.  

It should be noted, that this estimate is considered to independent of the previous discussion, since any effect arising from the variable $s$ is effectively canceled out by the use of equal but opposite pulses in these measurements and simulations.

\begin{figure}[h]
    \centering
    \includegraphics[width=85mm, keepaspectratio]{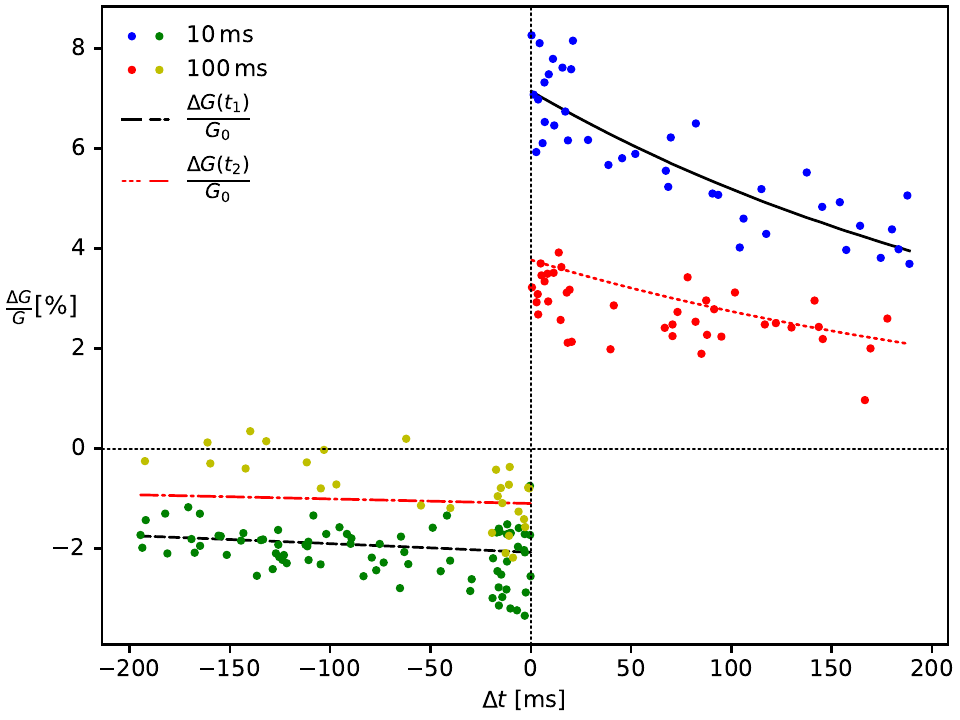}

    \caption{Fit and prediction of the STDP-like behavior. The dots show the experimentally measured relative conductance changes $\Delta G/G_0$ as a function of spike-timing difference $\Delta t$ for a paired-pulse protocol applied to the PCaPMA memristor.
    The black lines represent the model output  with parameter values fitted to the data at reading delay $t_1=10$\,ms. The red lines represent the prediction of the model, with the same parameters, when evaluated at reading delay $t_2=100$\,ms.}
    \label{fig:stdp}
\end{figure}

We calculate the difference in resting and STDP-induced conductance 
\[
\Delta G = \frac{\ker\ast dH \,(t_{1}) - g_0}{g_0}\, ,
\]
and estimate the parameters of the model to fit the measurements.
The results of the STDP-induced conductance at both sample times are shown in \cref{fig:stdp}. 
As a consequence, we verified that the decrease in STDP effects over time is a result of the previously modeled volatility mechanism.

\subsubsection*{Implementing saturation} 
Up to this point, we have ignored the effects of saturation in the modeling for
simplicity of presentation.  We thus implicitly assumed the effects
modeled so far lie in the linear regime of the saturation curve, so
that the subsequent addition of the saturation component will have
minimal effect.

The above discussed approach may be repeated with chosen coefficients
for the saturation module to obtain optimized model parameters that now
include the effects of saturation. Suitable coefficients for
$\sub{g}{max},\, \sub{g}{min},\, \sub{h}{max}$ and $\sub{h}{min}$ are
presented in the Appendix.

\section*{Methods}

\subsection*{Experimental data}

The proposed model is validated against experimental data obtained from
polymeric memristive devices that were synthesised and experimentally
characterized by~\citet{r:pfleger24}.  In low voltage regime, these
devices exhibit a rich set of synaptic-like behaviors, including
history-dependent conductance changes and volatile memory, which provide
the experimental grounding for our model's structure.

\subsection*{Polymeric memristor with neuro-synaptic properties}

The experimental validation of our model employs memristive devices
based on poly[N-(3-(9H-carbazol-9-yl)propyl)methacrylamide] (PCaPMA). In
these devices, a thin PCaPMA film is sandwiched between an indium tin
oxide (ITO) bottom electrode and a top electrode made of Au or Al. The
resulting structure functions as a one-port memristive element whose
conductance depends on the history of applied voltage pulses.

Panthi et al.\ demonstrated that such PCaPMA devices can emulate a broad
repertoire of synaptic functions that are central for neuromorphic
computing \citep{r:pfleger24}. When excited by trains of low-amplitude
voltage pulses, the device conductance can be gradually increased
(potentiation) or decreased (depression), depending on the polarity,
number, amplitude, and frequency of the pulses. The same platform
supports both short- and long-term plasticity, paired-pulse facilitation
and depression (PPF/D), spike-timing dependent plasticity (STDP), and
simple associative learning protocols, all on millisecond timescales.
These characteristics make the PCaPMA memristor a natural candidate for
validating compact models that aim to unify memristive switching,
volatility, and synaptic plasticity.

The microscopic origin of the resistive switching in PCaPMA has been
attributed to a combination of voltage-induced conformational changes of
the polymer, trapping and detrapping of charge carriers at localized
sites, and redox phenomena within the active layer
\citep{r:pfleger23,r:pfleger24}. A detailed microscopic description of
these intertwined processes is beyond the scope of the present work;
instead, we treat them as defining an effective internal state whose
dynamics is captured phenomenologically by the state variables of our
model and by the cumulative conductance function \(H(s,w)\).

Beyond the synaptic functions reported in \citet{r:pfleger24}, our
collaboration with the device-fabrication team provided access to
extensive relaxation and timing datasets obtained on the same class of
devices, including unpublished measurements. These data show that, after
potentiation by a finite pulse train, the conductance relaxes slowly
over several decades in time, with decay curves that are close to linear
when plotted in double-logarithmic coordinates (\Cref{fig:decay}).  This
long-tailed relaxation is inconsistent with a single exponential and
forms the empirical basis for our choice of a power-law-like memory
kernel in the volatility module.

For modeling purposes, the key device-level properties that we retain
from the PCaPMA system are therefore: (i) analog, gradual conductance
modulation by trains of voltage pulses; (ii) coexistence of volatile and
non-volatile plasticity on millisecond–second timescales; and (iii)
low-voltage operation compatible with mixed-signal neuromorphic
circuits. While our framework is formally material-agnostic, PCaPMA
provides a concrete polymer platform whose experimentally characterized
synaptic-like behavior and volatile relaxation impose quantitative
constraints on the kernel \( \mathrm{ker}(t) \), the cumulative
conductance \(H(s,w)\), and the saturation module.


\section*{Discussion}

We have introduced a modular memristor model that unifies a voltage-controlled memristive core, synaptic-like plasticity and volatile
relaxation within a single, computationally efficient framework.  By
grounding the modules for volatility and plasticity in principles from
condensed matter physics and computational neuroscience, respectively,
the model moves beyond pure phenomenology to provide a physically and
biologically plausible description of complex memristive behaviors.

\subsection*{Extending the state-of-the-art memristor models}

The current landscape of compact memristor modeling is dominated by
frameworks such as the TEAM and VTEAM models, which provide
computationally efficient descriptions of the core resistive switching
but are fundamentally phenomenological \citep{r:kvatinsky13,
r:kvatinsky15}. Other approaches have examined the dynamical aspects of
memristive switching, including shock-wave propagation phenomena that
govern commutation speed \citep{r:tang16prx}.  Our model extends this
state-of-the-art by incorporating two crucial, biologically inspired
features—synaptic plasticity and volatile memory—not as ad-hoc
additions, but as modules derived from established theoretical
principles.

In neuromorphic contexts, the STDP module enables biologically plausible
learning rules, facilitating simulations of associative memory and
adaptation \citep{r:jo10,r:zamarreno11, r:serrano13}.

To model synaptic plasticity, we implement a spike-timing-dependent
plasticity (STDP) rule inspired by the work of \citet{r:gerstner14}.
This approach is based on the concept of eligibility traces, a
cornerstone of modern theories of learning in the brain.  In this
framework, plasticity is a two-stage process. First, the
near-coincidence of pre- and post-synaptic activity creates a
short-lived synaptic "tag" or trace that marks the synapse as "eligible"
for modification.  Second, this temporary trace is only converted into a
lasting weight change if a third, modulatory signal—representing global
factors like reward, novelty, or surprise—arrives while the trace is
still active \citep{r:gerstner18}.  By adopting this mechanism, our
model moves beyond simple pairwise STDP to be extended toward and
flexible three-factor learning rules.  This capability aligns the model
with cutting-edge research in reinforcement learning and makes it a
compelling component for building truly adaptive neuromorphic systems.

To model memory volatility, we took inspiration from the theory of
linear viscoelasticity, which describes history-dependent phenomena in
materials science \citep{r:larson99}.  The mathematical structure is
that of a hereditary integral, where the current conductance state is a
convolution of the entire past history of voltage stimuli with a
decaying memory kernel.  On top of this linear convolution, we add a
final nonlinearity to shape the output conductance.  This structure is
analogous to the well-known linear-nonlinear  models widely used in
computational neuroscience to describe the relationship between a
stimulus and a neuron's firing rate \citep{r:gerstner14}.  In our model,
the linear filter captures the device's history-dependent memory, while
the subsequent nonlinear stage accounts for saturation and other
device-specific characteristics.

In contrast to the V-VTEAM model \citep{r:patni2024}, which implements a
stretched-exponential volatility model directly in the formulation of
the dynamics, our approach to volatility is threshold independent and
offers a broad range of possible decay characteristics.  For example, a
differential equation formulation for a $1/t$ kernel is notoriously
computationally unstable, but poses no difficulty in the convolution
approach.

\subsection*{Physical origin of the 1/t-like memory kernel}

The volatility module of our model is
governed by the hereditary kernel in \cref{eq:kernel} with an exponent
$\alpha \approx 0.03$ obtained from the log–log relaxation data in
\Cref{fig:decay-log}. The corresponding cumulative memory $K(t) =
\int_{-\infty}^{t} \ker(\tau)\,d\tau$ decays approximately as $K(t)\sim
1/t$ for large $t$.  In practice, the device therefore exhibits
effectively $1/t$-like ``loss of memory'' over time.

Power-law relaxations of this type are typical for transport in
disordered media. In a broad class of disordered conductors, charge
transport proceeds along percolating networks of microscopic pathways,
and charge carriers perform a continuous-time random walk (CTRW) with a
broad distribution of waiting times between hops
\citep{r:kirkpatrick73,r:scher73,r:stauffer94}. If the waiting-time
distribution $\psi(\tau)$ has a heavy tail, e.g.
$\psi(\tau)\sim\tau^{-1-\alpha}$ with $0<\alpha<1$, the resulting
transport is sub-diffusive and the macroscopic response to a
perturbation relaxes as a power law rather than as a sum of
exponentials. In our context, the ``response'' is the conductance of the
memristive device following a potentiating pulse train, and the observed
long-tailed decay directly motivates the $1/t$-like kernel in
\cref{eq:kernel}.

The viscoelastic analogy provides a
convenient heuristic explanation of this situation. In linear
viscoelasticity, a material with a broad spectrum of relaxation times
$\rho(\tau)$ can be represented as a superposition of many Maxwell
modes, and a spectrum $\rho(\tau)\propto\tau^{-1-\alpha}$ yields an
effective relaxation modulus $G(t)\propto t^{-\alpha}$
\citep{r:larson99,r:song23}. Our choice of
$\ker(t)\propto(t+\varepsilon)^{-(\alpha+1)}$ plays the same role: it
compactly encodes the presence of many internal relaxation processes
without having to resolve them individually at the level of the model.

For the PCaPMA devices used here, the microscopic origin of such a broad
relaxation spectrum is likely associated with structural and energetic
disorder in the polymer, together with charge trapping/detrapping and
redox processes in the active layer, as discussed in
\citet{r:pfleger24}. Different local environments and slow morphological
rearrangements would naturally generate a hierarchy of timescales. Our
model does not attempt to distinguish these mechanisms explicitly.
Instead, the 1/t-like kernel should be viewed as a phenomenological
summary of their cumulative effect, constrained by the experimentally
observed long-tailed relaxation in \Cref{fig:decay-log,fig:decay}.

Importantly, the argument leading to a power-law kernel is not specific
to PCaPMA. Percolation-controlled transport and CTRW-style dynamics are
common in organic and hybrid memristive systems \citep{r:wang20}. We
therefore expect that, when adapting the present framework to other
polymeric or soft-matter memristors, only the kernel exponent $\alpha$
and an overall time scale need to be adjusted, while the basic structure
of the volatility module -- a viscoelastic-type kernel acting on the
cumulative conductance function $H(s,w)$ -- remains unchanged.

\subsection*{Adapting model to polymeric memristor}

PCaPMA combines processability (solution casting/printing),
morphological stability (nm-scale roughness)\citep{r:pfleger23}, and low-voltage operation
with rich neurosynaptic functionality \citep{r:pfleger24}. Evidence from
transport and spectroscopy points to a multi-mechanism picture in which
conformational reorientation of carbazole side groups,
trapping/detrapping at localized states, interfacial barrier modulation,
and redox processes jointly modulate conductance; critically, repeated
pulsing produces analog, history-dependent updates consistent with
synaptic potentiation/depression at $\pm(0.3\text{--}0.6)\,$V and
$5\text{--}20\,$ms widths. Spike-rate dependence reveals an STP$\to$LTP
transition with repetition rate ($> 1$\,Hz), and paired-pulse and STDP
protocols yield causal/anti-causal weight changes with time constants in
the $\mathcal{O}(10^2)$\,ms range, compatible with our local-trace
plasticity module \citep{r:pfleger24,r:gerstner14}.

\subsection*{Prospective model extensions}

The modularity of the proposed model is one of its key strengths and
opens several avenues for future work. Each component—the non-volatile
core, the volatility kernel, and the plasticity rule—can be modified or
replaced independently to adapt the model to different physical devices
or to explore alternative theoretical assumptions. This allows for the
separate experimental characterization and fitting of each module,
simplifying the process of model validation and refinement for new
memristive technologies.

Furthermore, while the current model is deterministic, the experimental
data clearly exhibit stochasticity. This inherent randomness in
polymeric devices is not necessarily a flaw; it can be a functional
feature. For example, recent work has demonstrated that the stochastic
switching of polymeric memristors can be used as a physical entropy
source for applications in probabilistic computing \citep{r:foulger25}.
A natural extension of our work would be to incorporate a stochastic
component into the model, which would not only improve its descriptive
accuracy but also allow it to be used for designing and simulating such
novel computing paradigms.

The $1/t$ form is phenomenological; multiple microscopic routes
(trapping/detrapping spectra, filament reconfiguration, ionic drift) can
lead to similar envelopes. Future work should correlate $\alpha$ with
material properties (electrode, thickness, processing) and examine
frequency dependence.

\section*{Conclusions}

We have presented a modular and computationally efficient compact model
for memristors that exhibits both volatile memory and synaptic-like
plasticity.  The model's key innovation lies in the synthesis of several
modeling techniques that allow for memristor current-voltage
characteristics, synaptic-like plasticity effects and a general
description of relaxation dynamics with
saturation.  We have discussed how the volatile relaxation dynamics,
experimentally observed as a power-law decay, can be described by a
viscoelasticity-inspired hereditary integral whose form is physically
motivated by percolation theory.  Furthermore, we have demonstrated that
a biologically plausible model of STDP based on eligibility traces can
be implemented, endowing the future devices with a potential capacity
for advanced learning.

The complete model was shown to be in excellent quantitative agreement
with experimental data from polymeric memristors.  Furthermore, we believe
that the modularity and relative simplicity of the proposed model allow
its implementation (including validation) in a wide range of
memristive-like devices.  Therefore, this work provides a unified,
mathematically grounded framework for understanding and simulating
complex memristive devices, offering a valuable tool for advancing the
design of neuromorphic computing systems.


\subsection*{Funding}
This work was supported by The Czech Science Foundation project
No.~24-10384S, MEYS project Inter-Excellence No.~LUAUS24032 and the
Strategy AV\,21 Programme “Breakthrough Technologies for the Future --
Sensing, Digitisation, Artificial Intelligence, and Quantum
Technologies”. 

\subsection*{Author contributions}
D.H. and L.K. conceived and led the project, carried out the theoretical
development, implemented the model and simulations, analyzed the data,
and wrote the manuscript. S.H.F. and K.K. contributed to the theoretical
development and provided critical feedback and revisions to specific
sections of the manuscript. A.P., Y.R.P., J.P., and J.V. fabricated the
polymeric memristors and performed the experimental measurements. 

\subsection*{Competing interests}
The authors declare no competing interests.
All
authors reviewed the manuscript and approved the final version.

\subsection*{Data availability}

The source data of PPF/PPD and STDP can be accessed from
Zenodo at \url{https://zenodo.org/records/12685756}.

The source code for the model can be accessed at
\url{https://github.com/danielhabart/modular_memristor}

\appendix

\section*{Appendix: Numerical values}\label{numvals}

Here we collect numerical values of the parameters fitted in the
preceding sections.  They are not required for understanding the
structure of the model, but they provide a complete description of the
specific PCaPMA device and protocols used in our study. 

The first \Cref{tab:alpha} lists the kernel exponent $\alpha$.  The second
\cref{tab:coeff} summarizes the amplitudes $b$ and offsets $c$ of the
relaxation traces for each stimulation duration $T$, which determine the
functions $b(T)$ and $c(T)$ appearing in the cumulative conductance
function-reconstruction procedure \Cref{eq:g-trace}.  The cumulative
conductance function power law coefficient is shown in \Cref{tab:a}.
The \Cref{tab:Hs} lists cumulative conductance coefficients $h_{11}$ and
$h_{10}$, as well as model coefficients $\mu$ and $g_0$, estimated to fit
the potentiation segment of \Cref{fig:dh1}.

\Cref{tab:stdp} reports the parameters of the STDP module
(\Cref{stdpw}), which are used to generate the model STDP curve in
\Cref{fig:stdp}.  Our choice for the parameters of the saturation module
are presented in \Cref{tab:sat}.

Together, these values define a reproducible parameter set for the
particular device studied, while keeping the main text focused on the
general modeling framework.

\begin{table}[t]
\centering
\begin{tabular}{lc}
\\
\hline
$\alpha$ & 0.029 \\
\hline
\end{tabular}
\vspace{0.5em}
\caption{Volatility power law coefficient estimate.}
\label{tab:alpha}
\end{table}

\begin{table}[t]
\centering
\begin{tabular}[t]{lccccc}
     $T$ [s]    &     70   &    35   &    7    &  3.5  &  0.7 \\      
    \hline   
    \(b \)&     54.65&    41.44&    40.23&  33.07&  29.78\\
    \(c \)&     30.31&    30.43&    30.63&  30.98&  31.13\\
\hline
\end{tabular}
\vspace{0.5em}
\caption{Numerical values of the trace-specific amplitudes $b$ and offsets $c$ obtained from fitting the relaxation traces for different stimulation durations $T$.
These values determine the functions $b(T)$ and $c(T)$ in \cref{eq:g-trace}, which are used to derive the kernel function $\ker(t)$ and, in turn, the form of the cumulative conductance function $H(s)$ from the experimental data.}
\label{tab:coeff}
\end{table}

\begin{table}[t]
\centering
\begin{tabular}{lc}
\\
\hline
$a$ & -0.585 \\
\hline
\end{tabular}
\vspace{0.5em}
\caption{Cumulative conductance function power law coefficient estimate.}
\label{tab:a}
\end{table}

\begin{table}
    \centering
    \begin{tabular}{l|cccc}
         $T$ &  $h_{11}$    & $h_{10}$      & $\mu$ & $g_0$ \\
         \hline
         3.5 s &  $1.70\times10^{-6}$     &   $44.8\times10^{-6}$    &  $0.226\, V^{-1}$   & $0.307$ $\mu$S\\
         35 s & $2.78\times10^{-6}$    & $54.4\times10^{-6}$  &  $0.215 \, V^{-1}$ &  $0.308$ $\mu$S\\
    \end{tabular}
    \caption{Estimate of model parameters. Owing to the stochastic nature of the polymer-based memristor under study, we obtain a good fit per measured trace of potentiation stimulation.}\label{tab:Hs}
\end{table}
\begin{table}[t]
\centering
\begin{tabular}[t]{lcc}
    \hline
    $\sub{A}{+}$   & 10.9\\
    $\sub{A}{-}$   & 0.90 \\ 
    $\sub{\tau}{+}$   & 0.176 s\\
    $\sub{\tau}{-}$   & 0.619 s \\
    $\sub{h}{01}$ & $1.01\times10^{-4}$\\
    $\sub{h}{00}$ & $2.52\times10^{-3}$\\
\hline 
\end{tabular}
\vspace{0.5em}
\caption{Numerical values of the parameters of the STDP module. 
The parameters $\sub{A}{+}$, $\sub{A}{-}$ and $\sub{\tau}{+}$, $\sub{\tau}{-}$ define the shapes and time constants of the potentiating and depressing lobes in the local-variable STDP dynamics \Cref{stdpw}, the coefficients $\sub{h}{01}$ and $\sub{h}{00}$ determine the dependence of the cumulative conductance function $H$ on the variable $w$ as in \cref{eq:dH}.
These values are used to generate the model STDP curve shown in \Cref{fig:stdp} and to ensure consistency between the dynamical STDP formulation and the measured memristor response.}
\label{tab:stdp}
\end{table}

\begin{table}
\centering
\begin{tabular}[t]{lcc}
    \hline
    $\sub{g}{max}$   & $0.5$ $\mu$S\\
    $\sub{g}{min}$   & $0.117$ $\mu$S \\ 
\hline
\end{tabular}
\caption{The saturation module coefficients were chosen to minimize the effect of added saturation to the model.
The parameters $\sub{h}{max},\, \sub{h}{min}$ were set equal to $\sub{g}{max}$, $\sub{g}{min}$, respectively.}
\label{tab:sat}
\end{table}

\bibliography{references}

@article{r:chua71,
  title={Memristor—The missing circuit element},
  author={Chua, Leon O.},
  journal={IEEE Trans. Circuit Theory},
  volume={18},
  number={5},
  pages={507--519},
  year={1971}
}

@article{r:chua76,
  title={Memristive devices and systems},
  author={Chua, Leon O. and Kang, Sung-Mo},
  journal={Proc. IEEE},
  volume={64},
  number={2},
  pages={209--223},
  year={1976}
}

@article{r:strukov08,
  author = {Strukov, Dmitri B. and Snider, Gregory S. and Stewart, Duncan R. and Williams, R. Stanley},
  journal = {Nature},
  number = {7191},
  pages = {80--83},
  title = {The missing memristor found},
  volume = {453},
  year = {2008}
}

@article{r:yang13,
  title={Memristive devices for computing},
  author={Yang, J. Joshua and Strukov, Dmitri B. and Stewart, Duncan R.},
  journal={Nat. Nanotechnol.},
  volume={8},
  number={1},
  pages={13--24},
  year={2013}
}

@article{r:ielmini18,
  title={In-memory computing with resistive switching devices},
  author={Ielmini, Daniele and Wong, H-S Philip},
  journal={Nat. Electron.},
  volume={1},
  number={6},
  pages={333--343},
  year={2018}
}

@article{r:kvatinsky13,
  title={{TEAM}: ThrEshold Adaptive Memristor Model},
  author={Kvatinsky, Shahar and Friedman, Eby G. and Kolodny, Avinoam and Weiser, Uri C.},
  journal={IEEE Trans. Circuits Syst. I},
  volume={60},
  number={1},
  pages={211--221},
  year={2013}
}

@article{r:kvatinsky15,
  title={{VTEAM}: A General Model for Voltage-Controlled Memristors},
  author={Kvatinsky, Shahar and Ramadan, Nili and Friedman, Eby G. and Kolodny, Avinoam},
  journal={IEEE Trans. Circuits Syst. II},
  volume={62},
  number={8},
  pages={786--790},
  year={2015}
}

@article{r:wang17,
  title={Memristors with diffusive dynamics as synaptic emulators for neuromorphic computing},
  author={Wang, Zhongrui and Joshi, Saumil and Savel'ev, Sergey E. and Jiang, Hao and Midya, Rivu and Lin, Peng and Hu, Miao and Ge, Ning and Strachan, John Paul and Li, Zhiyong and Wu, Qing and Barnell, Mark and Li, Geng-Lin and Xin, Huolin L. and Williams, R. Stanley and Xia, Qiangfei and Yang, J. Joshua},
  journal={Nat. Mater.},
  volume={16},
  number={1},
  pages={101--108},
  year={2017}
}

@article{r:kim22,
  title={Prospects and applications of volatile memristors},
  author={Kim, Dahye and Jeon, Beomki and Lee, Yunseok and Kim, Doohyung and Cho, Youngboo and Kim, Sungjun},
  journal={Appl. Phys. Lett.},
  volume={121},
  number={1},
  pages={010501},
  year={2022}
}

@article{r:wang20,
  title={Recent advances of volatile memristors: Devices, mechanisms, and applications},
  author={Wang, Ruopeng and Yang, Jia-Qin and Mao, Jing-Yu and Wang, Zhan-Peng and Wu, Shuang and Zhou, Maojie and Chen, Tianyi and Zhou, Ye and Han, Su-Ting},
  journal={Adv. Intell. Syst.},
  volume={2},
  number={9},
  pages={2000055},
  year={2020}
}

@article{r:jo10,
  title={Nanoscale memristor device as synapse in neuromorphic systems},
  author={Jo, Sung Hyun and Chang, Ting and Ebong, Idongesit and Bhadviya, Bhavitavya B. and Mazumder, Pinaki and Lu, Wei},
  journal={Nano Lett.},
  volume={10},
  number={4},
  pages={1297--1301},
  year={2010}
}

@article{r:zamarreno11,
  title={On spike-timing-dependent-plasticity, memristive devices, and building a self-learning visual cortex},
  author={Zamarreño-Ramos, Carlos and Camuñas-Mesa, Luis A. and Pérez-Carrasco, José A. and Masquelier, Timothée and Serrano-Gotarredona, Teresa and Linares-Barranco, Bernabé},
  journal={Front. Neurosci.},
  volume={5},
  pages={26},
  year={2011}
}

@article{r:serrano13,
  title={STDP and STDP variations with memristors for spiking neuromorphic learning systems},
  author={Serrano-Gotarredona, Teresa and Masquelier, Timothée and Prodromakis, Themistoklis and Indiveri, Giacomo and Linares-Barranco, Bernabé},
  journal={Front. Neurosci.},
  volume={7},
  pages={2},
  year={2013}
}

@book{r:gerstner14,
  title={Neuronal Dynamics: From Single Neurons to Networks and Models of Cognition},
  author={Gerstner, Wulfram and Kistler, Werner M. and Naud, Richard and Paninski, Liam},
  publisher={Cambridge University Press},
  year={2014}
}

@article{r:pfleger23,
  author ="Panthi, Yadu Ram and Pfleger, Jiří and Výprachtický, Drahomír and Pandey, Ambika and Thottappali, Muhammed Arshad and Šeděnková, Ivana and Konefał, Magdalena and Foulger, Stephen H.",
  title  ="Rewritable resistive memory effect in poly[N-(3-(9H-carbazol-9-yl)propyl)-methacrylamide] memristor",
  journal  ="J. Mater. Chem. C",
  year  ="2023",
  volume  ="11",
  issue  ="48",
  pages  ="17093-17105",
  publisher  ="The Royal Society of Chemistry"
}

@article{r:pfleger24,
  title={Emulating synaptic plasticity with poly(N-(3-(9H-carbazol-9-yl)propyl)methacrylamide) memristor},
  author={Panthi, Yadu Ram and Pandey, Ambika and Šturcová, Adriana and V{\'y}prachtick{\'y}, Drahom{\'\i}r and Foulger, Stephen H. and Pfleger, Jiří},
  volume = 5,
  pages = 6388,
  journal={Mater. Adv.},
  year={2024}
}

@article{r:foulger21,
  title={Boolean and Elementary Algebra with a Roll-To-Roll Printed Electrochemical Memristor},
  author={Grant, Brayden and Bandera, Yuriy and Foulger, Stephen H. and Vilčáková, Jarmila and Sáha, Petr and Pfleger, Jiří},
  journal={Adv. Mater. Technol.},
  volume={7},
  number={5},
  pages={2101108},
  year={2022}
}

@unpublished{r:foulger24,
  title={Towards a hardware spiking neural network: Learning and adaptation with an environmentally sustainable polymer memristor},
  author={Foulger, Stephen H. and Bandera, Yuriy and Wanless, Travis and Luzinov, Igor and Cobb, Olivia and Sehorn, Michael G. and Kostal, Lubomir and Pfleger, Jiří and Vilčáková, Jarmila},
  note={submitted},
  year={2025}
}

@article{r:coleman61,
  title={Foundations of linear viscoelasticity},
  author={Coleman, B. D. and Noll, W.},
  journal={Rev. Mod. Phys.},
  volume={33},
  number={2},
  pages={239--249},
  year={1961}
}

@book{r:larson99,
  title={The structure and rheology of complex fluids},
  author={Larson, Ronald G.},
  publisher={Oxford University Press},
  year={1999}
}

@book{r:stauffer94,
  title={Introduction to Percolation Theory},
  author={Stauffer, Dietrich and Aharony, Ammon},
  publisher={Taylor \& Francis},
  year={1994}
}

@article{r:song23,
  title={Non-Maxwellian viscoelastic stress relaxations in soft matter},
  author={Song, Jake and Holten-Andersen, Niels and McKinley, Gareth H.},
  journal={Soft Matter},
  volume={19},
  number={41},
  pages={7885--7906},
  year={2023}
}

@article{r:foulger25,
  title   = {Polymeric Memristors as Entropy Sources for Probabilistic Bit Generation},
  author  = {Foulger, Stephen H. and Bandera, Yuriy and Luzinov, Igor and Wanless, Travis},
  journal = {Adv. Phys. Res.},
  year    = {2025},
  volume  = {4},
  pages   = {2400142}
}

@article{r:kirkpatrick73,
  title   = {Percolation and Conduction},
  author  = {Kirkpatrick, Scott},
  journal = {Rev. Mod. Phys.},
  volume  = {45},
  number  = {4},
  pages   = {574--588},
  year    = {1973}
}

@article{r:scher73,
  title   = {Stochastic Transport in a Disordered Solid. I. Theory},
  author  = {Scher, H. and Lax, M.},
  journal = {Phys. Rev. B},
  volume  = {7},
  number  = {10},
  pages   = {4491--4502},
  year    = {1973}
}

@article{r:gerstner18,
  title   = {Eligibility Traces and Plasticity on Behavioral Time Scales: Experimental Support of neoHebbian Three-Factor Learning Rules},
  author  = {Gerstner, Wulfram and Lehmann, Marco and Liakoni, Vasiliki and Corneil, Dane and Brea, Johanni},
  journal = {Front. Neural Circuits},
  volume  = {12},
  pages   = {53},
  year    = {2018}
}

@INPROCEEDINGS{r:patni2024,
  author={Patni, Tanay and Daniels, Rishona and Kvatinsky, Shahar},
  booktitle={2024 IEEE International Flexible Electronics Technology Conference (IFETC)}, 
  title={V-VTEAM: A Compact Behavioral Model for Volatile Memristors}, 
  year={2024},
  volume={},
  number={},
  pages={1-4}
}

@article{r:wu13prx,
  title   = {Nonvolatile Resistive Switching in Pt/LaAlO3/SrTiO3 Heterostructures},
  author  = {Wu, Shuxiang and Luo, Xin and Turner, Stuart and Peng, Haiyang and Lin, Weinan and Ding, Junfeng and David, Adrian and Wang, Biao and Van Tendeloo, Gustaaf and Wang, Junling and Wu, Tom},
  journal = {Phys. Rev. X},
  volume  = {3},
  number  = {4},
  pages   = {041027},
  year    = {2013}
  }

@article{r:tang16prx,
  title   = {Shock Waves and Commutation Speed of Memristors},
  author  = {Tang, Shao and Tesler, Federico and Gomez Marlasca, Fernando and Levy, Pablo and Dobrosavljevi{\'c}, V. and Rozenberg, Marcelo},
  journal = {Phys. Rev. X},
  volume  = {6},
  number  = {1},
  pages   = {011028},
  year    = {2016}
}

\end{document}